\documentclass[10pt,conference]{IEEEtran}
\usepackage{cite}
\usepackage{amsmath,amssymb,amsfonts}
\usepackage{algorithm}
\usepackage[noend]{algorithmic}

\usepackage{comment}
\usepackage{hyperref}
\usepackage{setspace}
\usepackage{caption}
\usepackage{subcaption}
\usepackage{graphicx}
\newcommand{\RNum}[1]{\uppercase\expandafter{\romannumeral #1\relax}}

\newcommand{\etal}{\textit{et al}. }

\def\BibTeX{{\rm B\kern-.05em{\sc i\kern-.025em b}\kern-.08em     T\kern-.1667em\lower.7ex\hbox{E}\kern-.125emX}}

\begin{document}
\title{Efficient Profit Maximization in Reliability Concerned Static Vehicular Cloud System}
\thispagestyle{plain}
\pagestyle{plain}

\author{\IEEEauthorblockN{Suvarthi Sarkar, Akshat Arun, Harshit Surekha, Aryabartta Sahu \textit{IEEE Senior Member}}
\IEEEauthorblockA{\textit{Dept. of CSE,  IIT Guwahati, Assam, India.} Emails:\{s.sarkar, a.arun, h.surekha, asahu\}@iitg.ac.in }
}

\setstretch{.95}

\maketitle

\begin{abstract}

Modern electric VUs are equipped with a variety of increasingly potent computing, communication, and storage resources, and with this tremendous computation power in their arsenal can be used to enhance the computing power of regular cloud systems, which is termed as vehicular cloud. Unlike in the traditional cloud computing resources, these vehicular cloud resource moves around and participates in the vehicular cloud for a sporadic duration at parking places, shopping malls, etc. This introduces the dynamic nature of vehicular resource participation in the vehicular cloud. As the user-submitted task gets allocated on these vehicular units for execution and the dynamic stay nature of vehicular units, enforce the system to ensure the reliability of task execution by allocating multiple redundant vehicular units for the task.

In this work, we are maximizing the profit of vehicular cloud by ensuring the reliability of task execution where user tasks come online manner with different revenue, execution, and deadline. We propose an efficient approach to solve this problem by considering (a) task classification based on the deadline and laxity of the task, (b) ordering of tasks for task admission based on the expected profit of the task, (c) classification of vehicular units based in expected residency time and reliability concerned redundant allocation of tasks of vehicular units considering this classification and (d) handing dynamic scenario of the vehicular unit leaving the cloud system by copying the maximum percentage of executed virtual machine of the task to the substitute unit. We compared our proposed profit maximization approach with the state of art approach and showed that our approach outperforms the state of art approach with an extra 10\% to 20\% profit margin.

\end{abstract}


\section{Introduction}

The processing requirements of applications from the user side are rising rapidly, this is due to the boom of real-time mobile applications, machine learning services, and Internet of Things services (such as augmented/virtual reality, interactive gaming, autonomous driving, e-health, etc.). Most of the time, the users or user devices connect to the internet, and users are subscribed to many cloud services \cite{Nesmachnow15, Manogaran21} and hence to speed up the user works, users outsource their computation workloads to distant cloud servers for task execution in the cloud computing system. Therefore, users are charged when they offload the tasks to the cloud system for faster execution. In such a system, a cloud system managing platform is typically needed for arranging the incoming tasks, accepting them, and handling as many computation demands from users as possible, so that the platform can increase its revenue. Hence the cloud service provider tries to maximize the profit from the users and provide good quality of service to the users \cite{Ngugen21,Chen11}.

On the other side of spectrum, the automotive sector is also undergoing a dramatic transition. People are being introduced to ``smart cars" and ``smart electric cars" on a regular basis these days. As reported on \cite{Soni22}, the autonomous vehicle industry is expected to grow remarkably, with its market value projected to increase from around USD 26.3 billion in 2021 to USD 56.3 billion by 2026. Similarly, in Naceur et al.\cite{Naceur16} observed a substantial rise in the number of autonomous vehicles, surging from 20 million to 70 million during the same period. Recently, there is an increasing number of car manufacturers, sellers, and owners have entered the autonomous vehicle industry. Typically autonomous vehicles have an onboard high-end computing facility, a lot of storage and other sensing devices that can sense the road and environment conditions to assure driving safety. As an illustration, autonomous vehicles are integrated with various hardware components, including GPUs like NVIDIA Tesla, GeForce GTX, Jetson, Tegra, among others, as well as FPGAs such as Xilinx Virtex, Intel Altera Stratix, and more \cite{Damaj21}. As a result, it makes sense to think such a vehicle as a ``computer on wheels". Vehicles onboard high-end computing resources are routinely underutilized, especially when parked. It is anticipated that using these vehicle computing resources to benefit society in a meaningful and productive way has a large and long-lasting effect. 

\begin{figure}[tb!]
    \centering
    \includegraphics[scale=1]{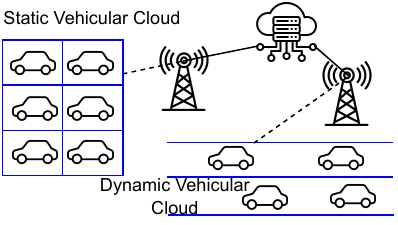}
    \caption{Vehicular Cloud Network \cite{Soni22}}
    \label{fig:VCN}
\end{figure}

The potent high-end computing resources of the smart autonomous car (vehicle or vehicular unit or car) can be used to enhance the computing power of the cloud system which is termed as the vehicular cloud (VC). We refer the participating cars as vehicular units (VUs). \autoref{fig:VCN} shows a typical vehicular cloud network, where the parked VUs in the parking lots can be part of the static vehicular cloud. These VUs augment their computation power to the cloud and enhance the computing capacity of the regular cloud. One the other hand, VUs in motion near the base station constitute a fast-dynamic vehicular cloud, where these moving VUs may connect with the vehicular cloud through the base station to the regular cloud, may seek computational assistance from the cloud. 

\begin{figure}[tb!]
    \centering
    \includegraphics[width=.47\textwidth]{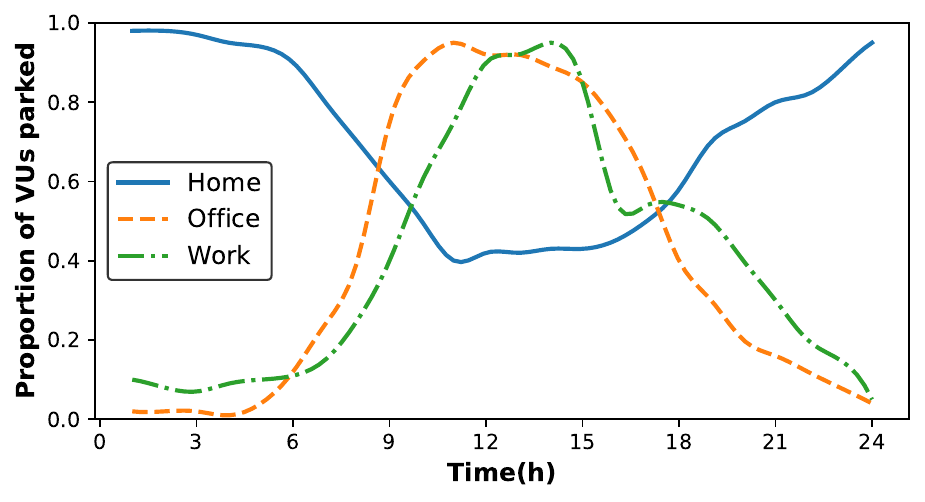}
    \caption{Parking Proportion Temporal Distribution \cite{Zhang14}}
    \label{fig:stay_p}
\end{figure}

In this study, since we focus only on the static vehicular cloud network (VCN), we simply refer it as vehicular cloud (VC). The VUs participate in the VC when they are parked in the parking lot. When a VU, such as a smart electric VU, leaves the parking lot, its connection to the VC is disconnected. More examples of static VC includes electronic VU charging stations, where electric VUs (EVs) spend an average of 7 hours charging \cite{Zhu14}. Another instance is an educational institution with residential area, where VUs remain parked or move within a limited boundary for extended periods. These VUs have the ability to connect to the VC, augmenting their computational capacity to the cloud system as most of the time the computational unit is unutilized. \autoref{fig:stay_p} illustrates the proportion of occupied parking spaces among home, office, and workplace categories. The parking facility at the office and workplace consistently experiences high occupancy rates during typical daytime office hours. Similarly, the home parking area exhibits full occupancy during night hours. Our work tries to take advantage of highly capable computation unit of VUs to facilitate user computation demands during their parking hours. 

Smart VUs have solid reasons to engage in the static VCN. Firstly, modern electric VUs are equipped with sophisticated high-end computing systems. The vehicular cloud awards the participating electric VUs some points, translating into monetary benefits as they lease their computation units. Secondly, the power consumption of the VU's computing system is significantly lower (approximately 5\%) compared to the power consumed during driving due to mechanical movement with high torque \cite{Pering06, energy11}. Thirdly, cloud-enabled VU parking lots are often equipped with charging stations, providing the convenience of recharging electric VUs if needed. Lastly, we assume that there exists sufficient security protocol to maintain the integrity and privacy of the VUs even if it lease out compute part of VU to the cloud.

The fact that vehicular clouds are quite dynamic is an important characteristic of a vehicular cloud; hence, it differs from traditional clouds. As VUs drive into the parking lot or VC area, more computing resources are made accessible and when VUs leave abruptly, they take their resources with them, developing an environment that is really very dynamic. In turn, the constantly fluctuating accessibility of computing resources as a result of VUs abruptly entering and leaving the VC causes an unstable computing environment, because of which reliability becomes a major issue. Let us imagine that a VU has a task scheduled on it, but the VU leaves the VC before the completion of the task. In this scenario, if specific safety precautions are not taken, the progress done by the VU is lost and must be redone, risking another VU leaving abruptly, and so on, until the task is eventually completed successfully. It is obvious that any potential loss of a VU's complete output due to an early departure must be minimized. Thus, not only maximizing the profit by accepting as many tasks as possible but also the reliability should be enhanced which could be done by executing a task on more than one VU in VC.

This study aims to enhance system reliability and maximize profit by ensuring timely task completion. The key contributions of this research are as follows:
\begin{itemize}
\item This work stands out by maximizing the profit of a vehicular cloud system while maintaining high reliability, a combination not found in existing literature. The proposed approach is also validated using real-life datasets.

\item In contrast to previous studies that assume uniform residency periods for all the VUs in parking lots, EV charging stations, or vehicular cloud networks, our research considers mixed residency times for VUs. This novel consideration enables a more realistic representation of the system.

\item Another significant contribution of this work is the development of an efficient approach that allows tasks to be executed redundantly, ensuring timely task completion and reliable service delivery, even in the face of potential failures or delays. By strategically leveraging VUs' diverse residency times, the system can enhance its overall performance and maintain a high level of reliability, making it a valuable addition to the field of vehicular cloud systems. We refer to this increased reliability as $MT99R$, which denote the time between two system failure having a probability of $0.01$.

\item The paper presents various strategies to optimize profit, such as task classification, task splitting, retry task completion even after failure, and improved task assignment strategy to gain an edge. These strategies are crucial in improving the overall efficiency and profitability of the vehicular cloud system.

 \end{itemize}

\section{Literature Review} \label{prevwork}

Due to its distinctive qualities, such as flexibility, elasticity, the availability of limitless processing resources, and a pay-as-you-use pricing model, cloud computing continues to gain popularity over time \cite{Li18,Nesmachnow15}. This prompts a lot of clients to move their operations to the cloud. Pay-as-you-use is a major component of cloud computing, which means customers must pay for the resources they use throughout the full usage period \cite{BUYYA09}. Chen \etal \cite{Chen11} suggested that in order to make customer satisfied, and abide by the Service level agreement (SLA) the system need to sacrifice a part of it's profit. On the other hand, the business goes in loss if the customer base is not satisfied. So, we proposed certain approaches where we design techniques where we tried to maximize the profit of the system and keeping the reliability above the threshold margin.

Different from traditional cloud networks, the vehicular cloud network is gaining its popularity \cite{Abuelela10,Basagni13,Eltoweissy10}. Due to its unique characteristics and usage, including standardization, effective traffic management, road safety, and infotainment, vehicular networking (VN) has become a prominent study subject. In 2010, Eltoweissy \etal \cite{Eltoweissy10} first introduced the concept of vehicular clouds (VCs), and defined it as ``a set of VUs whose corporate computing, sensing, communication, and physical resources may be coordinated and dynamically allocated to authorized users,". Abuelela \etal \cite{Abuelela10} suggested some  interesting, technical and cost effective implementations of VC. In Basagni \etal \cite{Basagni13}, they demonstrated how VCs may effectively address issues with dependability and availability, security and privacy, intelligent transportation systems, and similar issues.

As highlighted by Olariu and Florin \cite{Olariu17}, a significant portion of research in this domain tends to focus on a quantitative perspective rather than a qualitative one. Few studies conducted to explore the feasibility and reliability aspects in depth. In their work, Ghazizadeh \etal \cite{Ghazizadeh14} presented a novel approach using mixed integer linear programming to optimize efficiency and quality of service in Vehicular Clouds (VCs). However, a notable limitation of their approach is the assumption that VUs only enter and exit the network at specific checkpoints, which does not accurately reflect real-life scenarios where such constraints may not exist. Arif \etal \cite{Arif12} investigated data centers in airports, where vehicles have more extended residency times. They considered predetermined arrival and departure patterns, with vehicles typically staying for multiple days. Ensuring reliability in VCs with short and unpredictable vehicle residency times poses a significant challenge due to unforeseen events. One simplistic method to tackle this is to establish regular checkpoints at certain intervals, but \cite{Puyaphd} reported that such checkpoints often remain underutilized. On the other hand, Ghazizadeh \etal \cite{Ghazizadeh16} and Florin \etal \cite{Florin21} proposed an innovative approach involving redundant task execution to enhance system reliability in VCs. The key distinction between the two works lies in the assumption of vehicle residency time distribution, with the former considering a uniform distribution and the latter an exponential distribution. Furthermore, Yan \etal \cite{Yan13} conducted an extensive analysis of various security challenges in VCs and developed security schemes to address these issues.

We could not find any work which maximizes the profit of the system keeping the QoS above threshold in VCs.

Only a few of the existing works consider addressing these two issues simultaneously, even though they are both beneficial on their own for maximizing profit and improving dependability in the VC network. In this work, we aim to maximize profit in the vehicular network together with delivering reliability when carrying out duties to fill the gap in the existing literature.

\section{System Model and  Problem Formulation} \label{systemmodelPS}

\subsection{Vehicular Cloud  Environment}\label{VCE}

\begin{figure}[tb!]
    \centering
    \includegraphics[scale = .8]{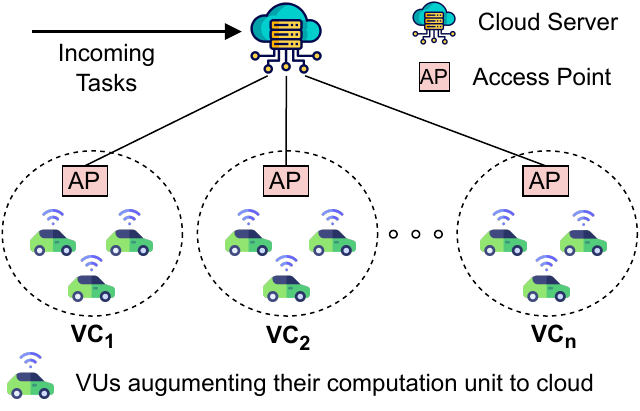}
    \caption{System Architecture - Availability Zone}
    \label{fig:architecture}
\end{figure}

We consider a vehicular cloud environment as shown in \autoref{fig:architecture}. It is basically a availability zone which is a collection of multiple VCs. The cloud server is logically one level higher at the center of the availability zone and is connected to multiple such VCs, and have a supply of incoming computational requests. We refer this computational requests as tasks. All the VCs provide computational support to the cloud to serve these tasks. Each VC is collection of many smart VUs in parking space. These VUs enhance the computing capacity of the VCs by leasing its compute resources to the VC. Each VU is connected to the VC via access point (AP) through a wired or wireless connection provided by the datacenter. The AP acts a router, the tasks pass through it and reach the VU. The AP also maintains a list for locating free VUs in the VC and assigning numerous VUs to a task for task execution. The VUs parked in the parking lot serve as compute resource of the datacenter. The statistic data shows that, during regular business hours, when the datacenter services are needed, the VC contains a sufficient number of VUs to make it possible to find a group of VUs that can be assigned to an incoming user task, even though VU residency times at parking lot is not directly available \cite{Zhang14}. 

\subsection{Vehicular Unit and Residency Time} 
We consider the VC that contains $M$ number of VUs, more specifically VUs at a given instant i.e. \(\textbf{C} = \{ C_1, C_2, \cdots, C_j, \cdots C_M \}\). In this section, we discuss about about the arrival and departure nature of the VUs and how the VUs execute the tasks.

\subsubsection{Residency and Arrival Time}
\label{RATime}
It would be beneficial to have knowledge of the approximate arrival time, departure time of the VUs parked at the parking lot, along with the total number of VUs. However, accurately predicting the exact time when VUs arrive or leave is nearly impossible. Instead, we can utilize arrival time and probability distribution function (PDF), which provides a probabilistic view of the VU's arrival time, as well as the residency time PDF to understand the distribution of residency time.
\begin{figure}[tb!]
    \centering
    \includegraphics[width=.5\textwidth]{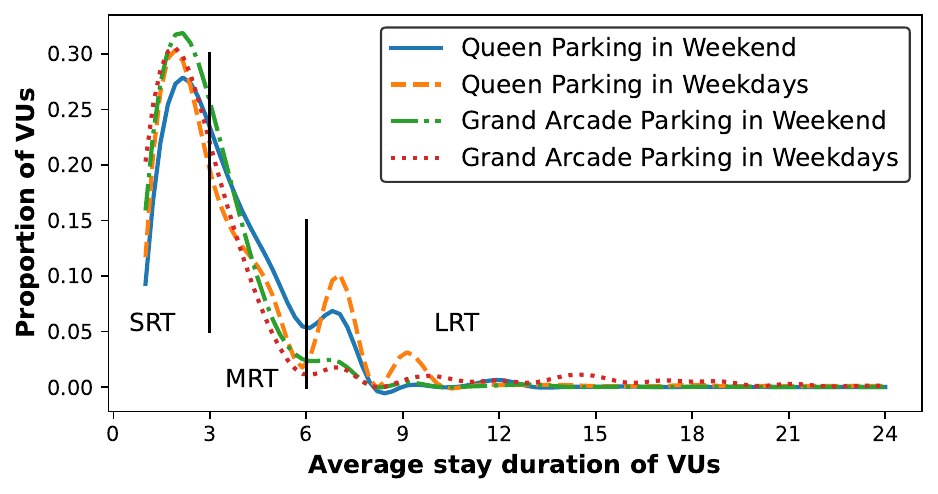}
    \caption{Parking duration of VUs in real-life dataset \cite{dataset_grand_arcade,dataset_queen_annes_terrace}}
    \label{fig:stay}
\end{figure}
The literature on vehicular networks commonly assumes that the Poisson process adequately captures the time of VUs' arrival at a parking facility \cite{Arif12}. Furthermore, an exponential distribution is often employed to estimate the residency times of VUs in a parking lot \cite{Florin21}. Based on this assumptions, we consider the residency times of VUs in the VC to be independent random variables, following an exponential distribution with parameter $\lambda$ . 

In \autoref{fig:stay}, the graph displays the mean parking duration for which all VUs parked at Queens Parking and Grand Arcade Parking between April 2012 and May 2020 \cite{dataset_grand_arcade,dataset_queen_annes_terrace}. Utilizing this information, we segment the pool of Vehicular Units (VUs) into three distinct categories:
\begin{itemize}
\item VUs with short residency time, referred to as Short Residency Time (SRT) VUs, with a mean residency time of \(t_{mrt}^s = 120 minutes\). In the real world, approximately 70\% of VUs belong to this group, and consists of all the VUs residing in the VC for less than 3 hours. 
\item VUs with medium-term residency time, termed as Medium Residency Time (MRT) VUs, with a mean residency time of \(t_{mrt}^m = 200 minutes\). About 20\% of VUs fall into this category, and they comprises of all the VUs having residency time between 3 to 6 hours. 
\item VUs with long residency time, denoted as Long Residency Time (LRT) VUs, with a mean residency time of \(t_{mrt}^l =400 minutes\). Approximately 10\% of VUs belong to this group, and they usually contains VUs parked for more than 6 hours. 
\end{itemize}

For smooth management of the parking lot, each VU specifies its intended stay duration, allowing the VC to be aware of the composition of VUs in different groups. Typically, VUs tend to stay around the same time they previously mentioned, but there is also a possibility of slight understay or overstay compared to their previously indicated time, ranging from minimal to significantly longer durations.

\subsubsection{Virtualization}
Virtualization serves as a fundamental foundation for Vehicular Clouds (VCs), enabling the deployment of Virtual Machines (VMs) on Vehicular Units (VUs) to deliver computation capabilities to the VC. We assume that each VU is equipped with a virtualizable on-board computational unit \cite{Florin21}. The Virtual Machine Monitor (VMM) is responsible for managing the mapping between the guest VM and the resources of the host VU. Within the guest VM, a guest operating system is hosted, running user applications to fulfill computation requests. To maintain simplicity, the assumption is made that each VU operates at most one VM. So, all the VUs are homogeneous and the VMs are of only one type.

Upon arrival at the VC, a VU requires a certain amount of time to configure the VM and guest OS. Once this configuration is completed, the VU becomes eligible to participate in task execution. It is important to note that if a VU enters the VC but leaves before its configuration is fully set up, that VU won't be considered for task execution. It's worth mentioning that the replication of the VM for redundant task execution and the subsequent transfer of the VM image occur relatively quickly compared to the overall task execution time. This is because the VM instance primarily contains metadata, which is typically a few megabytes in size. Moreover, the low transfer latency is facilitated by the close proximity of the VUs within the VC.

\subsection{Access Point (AP)}
As we have previously mentioned the AP maintains a list ($VU_{status}$) which contains the status of all the VUs in the VC. When a VU deaprts from the VC, the AP removes the VU from the list and when a  VU arrives, the AP appends it to the list. From the list, the AP can verify whether a VU is actively engaged in task execution or is available. The AP orders the VUs in order to the expected residency time left. Depending on the order and expected residency time left, the AP classifies the VUs into different categories of SRT, MRT and LRT. The AP updates this list a regular intervals.

In addition, the AP loads the requisite software onto the VUs to initiate task execution. After the selection of appropriate VUs for task execution, the AP also takes the responsibility of task migration to the appropriate VUs. The AP does continuous check-ins over the task execution after a brief interval to manage scenarios like VU leaving, task failure and completion.

\subsection{Task Environment}
\par{As stated earlier, users outsource their calculation workloads to distant cloud servers for task execution in the traditional cloud computing architecture. We have a set of $N$ independent tasks T = \(\{ \tau_1, \tau_2, \tau_3, ..., \tau_N \}\) with each task \(\tau_i\) is specified by a 4-tuple \(\{a_i, e_i, d_i, r_i\}\). Here, the tasks' respective arrival time, execution time, deadline, and revenue are represented by the numbers \(\{a, e, d, r\}\).} In our specific scenario, the execution times of tasks typically lasting a several hours. These tasks involve machine learning training models, where users commonly submit them to the cloud, and in return, they receive the trained model, which is essentially a collection of numerical values. We assume the user task execute serially on the VU but the redundant execution needs to be done on multiple VUs. Further, each task can execute in any one VC, the computation requirement of the task can not be divided among multiple VCs.

\subsection{\(T_n\) Task Assignment Strategy}
\label{jntaskass}
\par{We formulate our $T_n$ task assignment strategy by building upon the concepts discussed in the work of Florin \etal \cite{Florin21}. The objective of this approach is to execute a task on $n$ distinct VUs redundantly within the VC. The implementation of this redundant execution strategy guarantees a high level of task completion reliability.
The AP assigns $n$ VUs to each task who does the required computation in order to accomplish the successful execution till completion of the task. In the subsequent sections, we delve into diverse analyses and methods aimed at determining an appropriate value of $n$. Among these $n$ VUs, one is designated as the recruiter. In addition to providing computation for the task execution, the recruiter also assumes the supplementary responsibility of identifying a replacement VU in case any of the current selected VU leaves the VC.}

\par{There are two possible outcomes in this situation:
\begin{enumerate}
\item The VU leaving the VC is not a recruiter. 
\item The VU leaving the VC is a recruiter. 
\end{enumerate} 
}

\par{In the first case where a VU that is departing the VC is not a recruiter. The special work of the recruiter is to maintain the number of VUs assigned for a task ensuring reliability during execution of the assigned task. The recruiter starts the hiring process by saving the status of its guest VM that has completed the most work amongst the VUs executing that task. Let $k$ denote the number of departed VUs, which is updated if any of the remaining VUs leave while recruitment procedure is still in progress. A system failure is notified if $k = n$. Otherwise, a fresh recruiting effort is launched as previously said. Once a recruiting operation is underway, this process is repeated until no VUs leaves the VC during recruitment. 

The AP now assigns $k$ fresh VUs from the parking lot's pool of available VUs and broadcasts the most updated instance of VM image to them. The time it takes to successfully replace the departed VU with a new VU is $t_{mttr}$. With this, there are now $n$ VUs again, actively engaged in the user's task. Now let us consider the second instance, where a VU departing the VC is a recruiter. The AP selects a new recruiter amongst the remaining VUs and the steps as mentioned for the first instance are executed.}

\begin{figure}[tb!]
    \centering
    \includegraphics[scale=0.75]{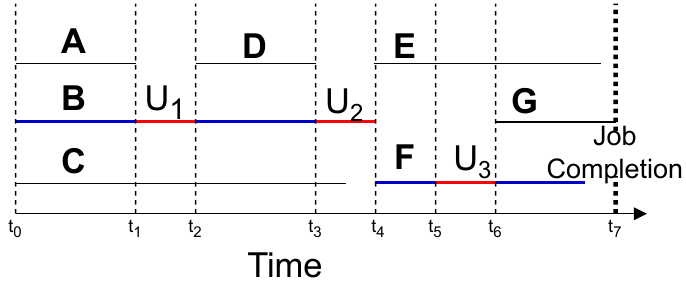}
    \caption{Illustrating the \(T_n\) task assignment strategy for $n=3$}
    \label{fig:Jn}
\end{figure}

Consider the scenario show in \autoref{fig:Jn}, where $3$ VUs are assigned to execute a user task. Suppose on VUs A, B, and C, the user task is allocated to execute at time $t_0$ and the VU B is designated as the recruiter. Suppose A exits the VC at time $t_1$, B discovers that VU A has completed the same amount of work (30\%) as the other VUs up to that point. B starts to making a copy of its own VM's state as everyone have done maximum work. Once it has saved its VM state, it selects a VU D, transfers the saved virtual machine to the  VU D, and then resumes the task on the VU D at time $t_2$, this process of recruiting new VU and starting execution of copies of task in the newly recruited VU takes some time $U_1$ (say mean recruitment time or mean replacement time denoted by $t_{mttr}$). The VU C keeps executing the task in the meantime. Suppose D departs from the VC at time $t_3$. Again the VU B discovers that every VU has completed equal work (suppose 50\%). B starts making a copy of its own VM state. Again, suppose the VU C departs as B is saving the state of its VM between time $t_3$ and $t_4$. Then the B must first recruit two VUs, E and F, copy the saved VM to both, and then resume the task on E and F after saving is complete (at time $t_4$). Now, suppose the VU B  departs from the VC, VU F is chosen randomly by the access point to be the recruiter. Now the VU F determines that B has completed 80\% of the task up till now, which is the maximum amount, so F recruits VU G, copy the VM state of B to the VU G. After the operation, G resumes task execution. F also overwrites its VM state with most updated VM state of B. As shown in the above figure, one of E, F, or G eventually completes the task and informs the AP of the outcome. The access point stops the task execution of rest of VUs, in order to make the best use of the resources. As shown in \autoref{fig:Jn}, the replacement time are $U_1$, $U_2$ and $U_3$, and we say mean replacement time ($t_{mttr}$) is average of all the replacement time. The replacement time includes the recruitment time, copying the VM to target new VU and starting execution task in the new VU.

Given our focus on tasks with significantly longer execution times, the replication time and VM instance transfer time are considerably shorter compared tp the task execution time. In our approach, we contribute by updating the newly recruited VUs with the VM instance with maximum task execution, which was not taken into account in the prior work \cite{Florin21}. In the previous approach, the newly recruited VU received the old VM instance from the VU that had left.

Although the VMs on VUs are homogeneous, the execution speed may vary due to several factors like, network failure, memory and compute unit availability. This is the reason why task execution speed varies in VMs.
\subsection{Cost Model}
The actual cost spent by the service provider to execute task $\tau_i$ with $n$ VUs for $e_i$ amount of time can be defined as:
\begin{equation}\label{eqn:cost}
     cost_i =  K * n_i * e_i 
\end{equation}
where $K$ is the actual cost spent per unit VU per unit time, \(n_i\) is the number of VUs assigned for executing task \(\tau_i\) and \(e_i\) is the execution time of the task \(\tau_i\).

\subsection{Problem Statement}

In order to optimize total profit, we want to schedule a set of $N$ tasks on homogeneous VMs that are present inside the VU. We call this problem to be Constraint Aware Profit Maximization (CAPM) problem. The CAPM problem can be defined as follows:
\begin{equation}\label{}
    Maximize \sum_{i}^{N} profit_i * x_i
\end{equation}
where $x_i$ is a indicator variable, $x_i=1$ signifies that the $\tau_i$ is completed by the availability zone within $d_i$, $x_i=0$ signifies the opposite. \(profit_i\) is the profit of executing a task \(\tau_i\) and can be written as:
\begin{equation}\label{}
profit_i = {r_i - cost_i}    
\end{equation}
where \(cost_i\) can be calculated from the \autoref{eqn:cost} and \(r_i\) is the amount earned by the system when a task is completed successfully. If a task fails, when all the VUs involved in the task's execution depart the vehicular cloud, the task is redone using a fresh group of VUs, and if the task is finished beyond the deadline, a penalty is incurred. While the primary goal is to maximize VC profits, it's equally imperative to maintain task execution reliability above a predefined threshold. In the next section, we delve into the strategies employed by the VC to guarantee the requisite level of reliability for incoming tasks.

\begin{table}[tb!]
\footnotesize
\centering
\begin{tabular}{ | l | l | } 
  \hline
  \textbf{Notation} & \textbf{Definition}  \\
  \hline

  \ T & Set of tasks \{${\tau_1,\tau_2,\cdots\,\tau_N}$\} \\
  $N$ & Number of tasks in Availability Zone \\
  $M$ & Number of VUs in a VC \\
  $t_{num}$ & Number of tasks executing in a VC \\
  $T_{t_{num}}$ & Set of running task in VC \\
  \ \(a_i, e_i, d_i, r_i\) & Arrival time, expected execution time, deadline \ \\ & and revenue earned of task \(\tau_i\)\\ 
  \ \(e'_i\) & Expected execution time left of task \(\tau_i\)\\
  \ \(l_i\) & Laxity of task \(\tau_i\)\\
  \ \(n_{s_i}\) & No. of small VUs to execute task \(\tau_i\)\\
  \ \(n_{m_i}\) & No. of medium VUs to execute task \(\tau_i\)\\
  \ \(n_{l_i}\) & No. of large VUs to execute task \(\tau_i\)\\
  \ \(t_{mrt}^s\), \ \(t_{mrt}^m\), \ \(t_{mrt}^l\) & Mean residency time of SRT VU (120 min),
  \\ & MRT VU (200 min) and LRT VU (400 min)\\ 
  \ \(cost_i\) & Actual cost incurred by VC to executute task \(\tau_i\)\\
  \ \(MTTF_{T_n}\) & Mean time to failure when $n$ \\ & VUs is redundantly executing a task \\
  \ \(MT99R\) & Mean time for 99\% reliability\\
  \ \(t_{mttr}\) & Time to replace a departed VU with another VU  \\
  $t_{mrt}$ & Mean residency time or sometimes mean \\
  \ \(profit_i\) & Profit gained by running task \(\tau_i\)\\
  \ \(g_i\), $x_i$, $VU\_type_i$ & Num. of checkpoint, redundancy, VU type of \(\tau_i\)\\
  \ \(n_{VU}^i\) & Number of VUs allocated to \(\tau_i\) in each group\\
  $Wdone_{VU_j}$ & Workdone by $VU_j$ \\
  \ \(n^i_{left}\) & Number of VUs leave the VC while executing \(\tau_i\)\\
 \ \(n_i\) & Number of VUs still executing \(\tau_i\)\\
  \ \(VU_R^i\) & Recruiter of \(\tau_i\) in a group\\
  $n_{min}^{s}$, $n_{min}^{m}$, $n_{min}^{l}$ & Minimum number of VUs in a group for SRT,   \\
   & MRT and LRT VUs for redundant execution \\  
  $LUT$ & Lookup table \\   
  $n^l_{VU}$ & number of idle LRT VUs present in a VC\\
  $T_{VU}$ & Reserved LRT VUs for critical tasks in each VC\\
  \hline
\end{tabular}
\caption{\label{Table 1:}Notations used}
\end{table}

\subsection{System Reliability}
\subsubsection{State of the Art Method of Calculation Reliability}

\begin{figure}[tb!]
    \centering
    \includegraphics[scale=1]{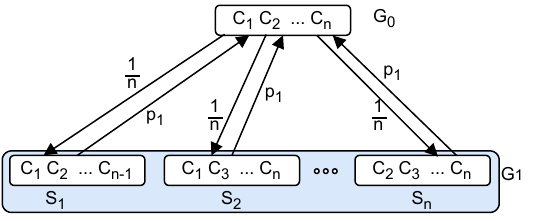}
    \caption{Illustrating groups $G_0$ and $G_1$  for user task $\tau_i$\cite{Florin21}}
    \label{fig:basepaper}
\end{figure}

In the study conducted by Florin \etal \cite{Florin21}, system reliability is characterized as $MTTF_{T_{n}}$. The term $MTTF_{T_{n}}$ denotes the duration from the initiation of task execution to the point where the probability of system failure reaches 0.5, where $n$ VUs are redundantly executing $\tau_i$. The MTTF of the $T_n$ approach is modelled using a semi-Markov process. This process simplifies states into groups based on the number of VUs assigned to a task. As shown in \autoref{fig:basepaper}, adopted from Florin \etal \cite{Florin21}, group \(G_0\) comprises of single state ($S_0$), where VUs \(C_1\), \(C_2\) through \(C_n\) denote the $n$ VUs that are allocated to user's task $\tau_i$. Group \(G_1\) comprises $n$ states denoted as $S_1$, $S_2$, $\cdots S_n$. Each state have $n-1$ VUs, assigned for the execution of  task $\tau_i$. It's worth noting that these states in same group, while all having the same number of VUs, vary from one another in terms of which specific VUs are present in the VC. Similarly, there are $n-2$ states in $G_2$ denoted as $S_{1,2}, S_{1,3}, \cdots S_{n-1,n}$. The indices at the bottom of the states represents which VUs have left the VC.

\begin{figure}[tb!]
    \centering
    \includegraphics[scale=0.45]{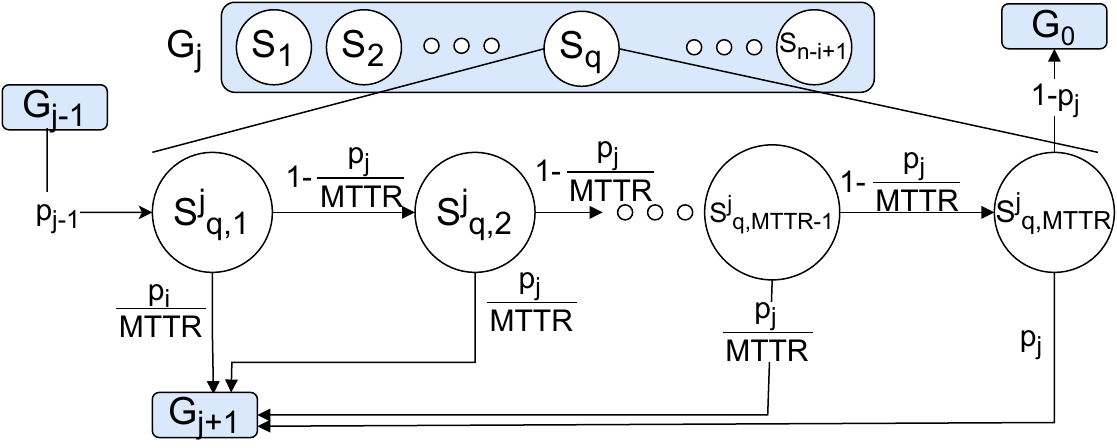}
    \caption{Decomposing of State $S_i$ to intermediate states $S_{i,j}$ }
    \label{fig:MC2}
\end{figure}

Transition from state $S_0$ to any state of group $G_1$ occurs when a VU assigned to the execution of task $\tau_i$ leaves the VC. The transition depends on the departing VU. Notably, the probability of a transition from the state of \(G_0\) (i.e. state $S_0$) to one of the states in group \(G_1\) is $\frac{1}{n}$. We assume that  all VUs leave the VC with equal probability. Similarly, the group \(G_j\) consists of states corresponding to $n-j$ VUs that are allocated to the task $\tau_i$ execution as shown in \autoref{fig:MC2}. There are $n-j+1$ states in group $G_j$ depending on which VU has departed the VC. Conclusively, $j$  signifies the count of VUs that has departed the VC while executing $\tau_i$ of the assigned $n$ VUs. This is because the system settles to the same state irrespective of order of leaving of particular VUs. For example, let us consider two case where $C_k$ and $C_l$ leaves the VC when the system is in state $S_0$. The system goes from $S_0$ to $S_k$ and $S_l$ respectively. The case of $C_k$ leaving the VC is followed by $C_l$, and $C_l$ leaving is followed by $C_k$ leaving. In either case, the system goes to state $S_{k,l}$ from $S_k$ and $S_l$ respectively. 

\begin{figure}[tb!]
    \centering
    \includegraphics[scale=0.65]{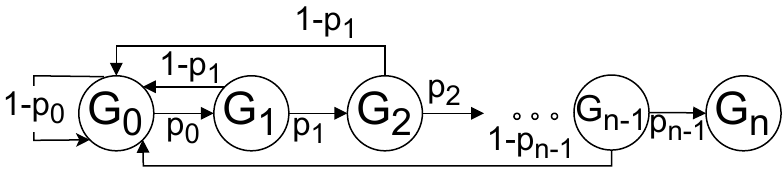}
    \caption{Proposed Markov Chain for Reliability Calculation}
    \label{fig:MC}
\end{figure}

Each state in group $G_j$ has the probability of $1-p_j$ to go state $S_0$, and a probability of $p_j$ to move from any state of group $G_j$ to any state in $G_{j+1}, \forall j < n $ as shown in \autoref{fig:MC}. Transition from $G_j$ to $G_0$ is feasible through simultaneous recruitment. The one state ($S_n$) in group $G_n$ denotes the state when every VU has departed from the VC assigned to the task. This state represents the system failure for the task.

Suppose the system is at any state of group $G_j$. If the recruiting effort in that state of group $G_j$ is successful and $j$ VUs are recruited, the system goes to $S_0$. As proved by Florin \etal \cite{Florin21}, value of \(1-p_i\) is represented in \autoref{eqn:e1}. 
\begin{equation}\label{eqn:e1}
     1-p_j = \frac{\mu}{\mu + \lambda (n-j)}
\end{equation}
where, \(\lambda\) represents the parameter of the exponential distribution for VU's residency time in the VC, and \(\mu\) signifies the parameter of the exponential distribution for VU recruitment time.\\

\subsubsection{Proposed Method for reliability calculation}
Now, let's define a Markov chain for the reliability equations. Suppose for a task $\tau_i$, $n$ VUs are assigned. 

Our objective is to compute the failure probability: the likelihood of transitioning from a state in $G_0$ to the state in group $G_n$. Let $Er_j$ denote the probability to reach state in group $G_n$ from some state in group $G_j$ i.e. probability of system failure from group $G_j$.
\begin{equation}\label{}
    Er_j = p_j*Er_{j+1} + (1-p_j)*Er_0
\end{equation}

After simplifying the equations we get, $Er_0 = Er_1 = ... = Er_n$. Since $G_n$ is the final state, we see that $Er_n$ is always be 1. So the probability to arrive at $G_n$ from $G_0$ is 1, if the task runs for $\infty$ unit of time. This makes sense also because there are no outwards arrow from $G_n$ and hence given sufficient infinite time the system always land up in $G_n$ no matter from where they start. However, the challenge is calculating the failure probability within a specific time frame. As noted by Florin \etal \cite{Florin21}  while they provided the expected stay time in any state of group $G_j$, address the probability distribution of the system being in different groups after a certain interval. This is important because we do not want the system to reach $G_n$ during the execution of any task.

To address this, we begin with a simplified scenario where the average stay period in any state of group $G_j$ be $t_j=1$ $\forall i$. Let $\hat{A}$ denote a matrix where $\hat{A}_{j,k}$ denotes the probability of directly landing in $G_k$ from $G_j$.
\begin{equation}
\hat{A}_{j,k}=
    \begin{cases}
        p_j & \text{if } k = j+1\\
        1-p_j & \text{if } k = 0\\
        0 & \text{otherwise}
    \end{cases}
\end{equation}

Let $\hat{A} = A^{\alpha}$ then $\hat{A}_{j,k}$ denotes the probability to land in any state of group $G_k$ from any state of group $G_j$ in exactly $\alpha$ jumps. So if $t_j=1$ $\forall j$ then we can interpret $\hat{A}_{j,k}$ as the probability to land in any state of group $G_k$ from any state of group $G_j$ after exactly $\alpha$ time unit. Now we can extend this definition for some constant $t_j$, i.e. if $t_j=\beta$ $\forall j$ then the probability to land in $S_k$ from $S_j$ after exactly $\alpha$ time unit is given by $A^{\lceil \frac{\alpha}{\beta} \rceil}_{j,k}$.

Now let us consider the final and the most generalised case that is $t_j$ can be different for all of the states of groups. To handle this case we have to follow an entire new strategy. Suppose for some state $S_q$ in group $G_i$ the average stay period of that state is $t_i$ then we decompose this state into several substates each with an stay period of $1$ unit time. Now the number of maximum stay period in any state in any group (other than $G_0$ and $G_n$) should not exceed $t_{mttr}$ unit of time. This is because once the system goes to any other state it tries to go to $G_0$ unless in case of task failure ($G_n$). So the number of substates is $t_{mttr}$. When the system goes to state $S_q$ from any states in group $G_{j-1}$, it goes to initial substate of $S_{q,1}^j$. The system moves to $S^j_{q,k}$ after $k-1$ unit of time if none of the VUs leave the VC during that time interval. In case another VU leaves the VC, then system goes to a state in group $G_{j+1}$ with probability $\frac{p_j}{mttr}$. The system can go to substate $S^j_{q,2}$ from $S^j_{q,1}$ with probability $1-\frac{p_j}{mttr}$, similarly from $S^j_{q,2}$ the system can go to $S^j_{q,3}$ with same probability and so on. From the last substate $S^j_{q,mttr}$, the system can go to $G_0$ with probability $1-p_j$ or to any state in $G_{j+1}$ with probability $p_j$. 

\subsubsection{Analysis}
\begin{figure}[tb!]
    \centering
    \includegraphics[width=0.5\textwidth]{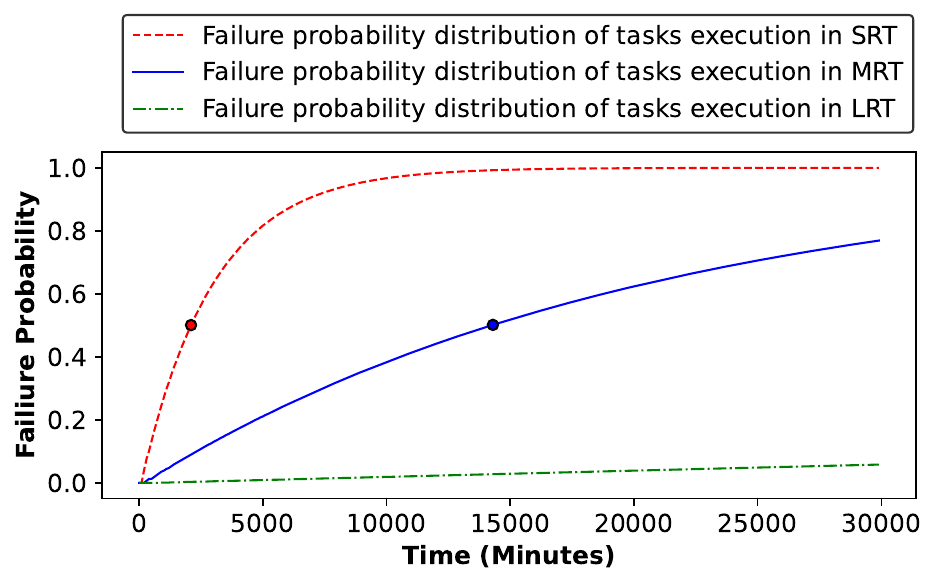}
    \caption{CDF of failure Probability is plotted against time. The $MTTF_{T_3}$ for SRT and MRT VUs is marked.}
    \label{fig:probab}
\end{figure}

Using the aforementioned process, we plotted the result of cumulative distribution function (CDF) failure probability with time as shown in \autoref{fig:probab}. The figure illustrates the change in CDF of failure probability over time for SRT, MRT and LRT VUs. \cite{Florin21} proposed a term $MTTF_{T_3}$, denoting the time at which the probability of task failure up to or before that specific point is $0.5$. Here, $3$ signifies the count of redundant task executions. The graph showcases $MTTF_{T_3}$ for SRT and MRT VUs, which is approximately 2200 and 14000 minutes, respectively. For LRT, $MTTF_{T_3}$ stands at roughly 117200 minutes. 

\begin{figure}[tb!]
    \centering
    \includegraphics[width=0.5\textwidth]{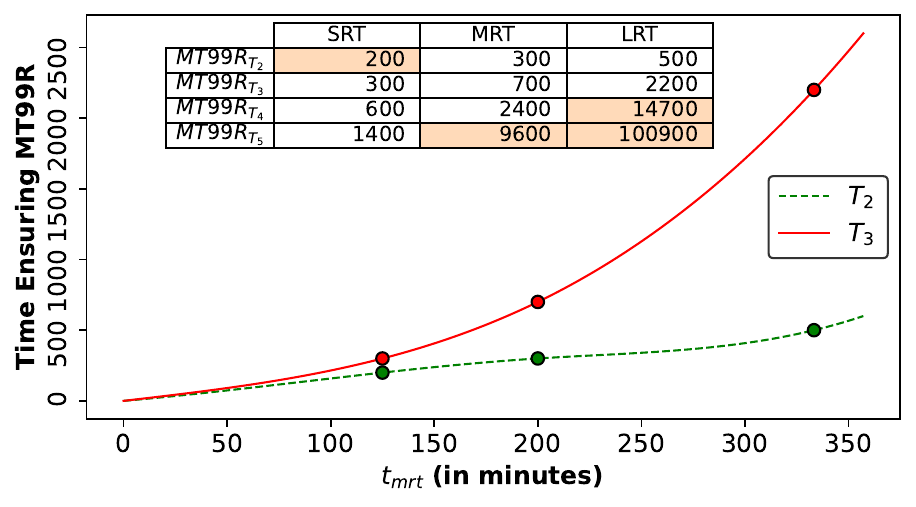}
    \caption{$MT99R_{T_3}$ and $MT99R_{T_2}$ is plotted against $t_{mrt}$ of VUs. The $t_{mrt}$ for SRT, MRT and LRT VUs are marked left to right. The table represents $MT99R_{T_2}$, $MT99R_{T_3}$, $MT99R_{T_4}$, $MT99R_{T_5}$ for SRT, MRT and LRT VUs. Some values are marked in different colour because we have not used that combination in our proposed approach.}
    \label{fig:MT99R}
\end{figure}

In the approach proposed by Florin et al. \cite{Florin21} the task completion reliability reaches only 50\%. This level of reliability is unacceptable for a practical system, as traditional edge-cloud systems offer a reliability of 99.99\%. Transitioning to a 99\% reliability is acceptable, given the significant cost reduction, but a shift to 50\% reliability is infeasible. We propose a term Mean Time with 99\% Reliability ($MT99R$), which denotes the time at which the probability of task failure up to or before that point is $0.01$. \autoref{fig:MT99R} shows $MT99R_{T_2}$, $MT99R_{T_3}$, $MT99R_{T_4}$, $MT99R_{T_5}$ for SRT, MRT and LRT VUs. $MT99R_{T_2}$ for SRT, $MT99R_{T_5}$ for MRT and $MT99R_{T_4}$, $MT99R_{T_5}$ for LRT are marked in different colour, because we do not use these values in task allocation in our approach. We provide a more detailed explaination in \autoref{split_long_short_tasks}.

The result of this graph is stored in a lookup table $LUT$, which is available to the AP. The AP refers to this lookup table to determine the appropriate number of VUs allocated to a task and number of task splits, depending on the type, count and execution time of the task. 

\section{Overall Task Scheduling Approach}\label{SolutionApp}
\label{SolutionApp}
\begin{figure}[tb!]
    \centering
    \includegraphics[scale=1.05]{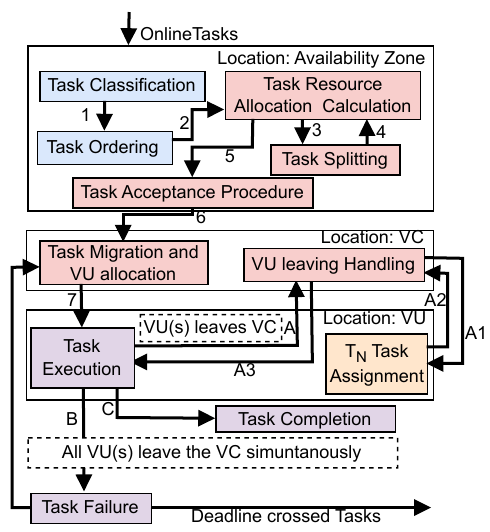}
    \caption{The figure shows the system flowchart of the proposed approach.The submodules coloured in blue, red and yellow is initiated by Availability Zone, AP of respective VC, Recruiter respectively. The different system states are marked in violet, and different event are marked with dotted boxes. The submodules are ordered in numerical order as the order of their occurrences.}
    \label{fig:fc}
\end{figure}

The overall  flow of solution approach is shown in the \autoref{fig:fc}. When a user task arrives to the availability zone, then the availability zone classifies and orders the tasks depending on some approach. Using this ordering the AP of VC chooses some tasks if it can accommodate the task with sufficient VUs for reliable ensured execution. The AP also supervises over the task execution to handle scenarios like task migration, VU allocation, VU leaving, task failure and task completion.

The overall solution approach starts when the availability zone has a pool of tasks. The approach has nine sub modules and these are (a) Task Classification, (b) Task Ordering, (c) Task Splitting, (d) VU allocation calculation, (e)Task Acceptance, (f) VU Leaving Handling, (g) $T_n$ Task Assignment, (h) Task Failure, (i) Task Completion. The following subsections describes about these sub modules in details.

\subsection{Task Classification}

\par{Based on characteristics of user tasks, the availability zone classifies the
user tasks into two classes: (a)  critical tasks and (b) non-critical tasks.  A task $\tau_i$ is considered to be critical if one of the two conditions satisfy: (a) Laxity ($l_i=d_i-a_i-e_i$) of the task is less than (1 + $\delta$) times of its execution time ($e_i$), where $\delta$ is a constant factor. We consider $\delta = 0.1$. (b) its execution time falls under a threshold limit $e_{threshold\_critical}$. The time complexity of this classification is \(O(N)\), where $N$ is the total number of task at any instant.}

\subsection{Task Ordering}
\label{ss:to}
Ordering and selecting the tasks in proper order is pivotal for maximizing the profit of the VCs. 
A VC enhances its pocket if higher number of highly profitable tasks are completed faster than expected. When incoming tasks reach the availability zone, the availability zone orders tasks using various methods outlined below.  Subsequently, AP of VCs utilize this sequence to select tasks available at first position if they can provide allocate sufficient number of VUs, else the APs move to the next task in the sequence.  Upon the selection of a task, the subsequent task in line moves up to occupy the previous position in sequence. Here, we explore some heuristics based approaches in order to analyze how they affect the maximization of profit. The considered task ordering heuristics are as follows: 
\begin{itemize}

\item Expected profit (EP): Tasks with greater EPs are thought to be more successful. In this case, the tasks are arranged in descending order of the predicted profit.
\item Revenue (RV): In this case, tasks that have higher revenue are considered more profitable. In this case, the tasks are arranged in decreasing order of the revenue ($r_i$).
\item Revenue per unit of execution time (RPE): In this case, tasks are arranged according to their revenue per unit execution time (RPE), which is listed in descending order. The RPE of the task $t_i$ is $\frac{r_i}{e_i}$
\item Earliest Due Date (EDD): Using this criterion, tasks are arranged in increasing order of deadline of the tasks, it is also called the due date ordering.
\item First Come, First Served (FCFS): Tasks are arranged in ascending order based on their arrival times.
\item GUS approach:  In this scenario, tasks are organized in ascending order of the fraction of laxity of a task to its execution time, denoted as $\frac{l_i}{e_i}$. 
\end{itemize}

The time complexity of this ordering is \(O(N logN)\), where $N$ is the total number of task at any instant.

\subsection{Splitting of Longer Tasks into Shorter Tasks} \label{split_long_short_tasks}

Splitting longer tasks into shorter tasks often proves advantageous in terms of reducing VU computation wastage and cost effective. Consider a scenario where a task requires 2100 minutes of execution time. From \autoref{fig:MT99R} we can observe that it requires $2100/300 = 7$ checkpoints if each group contains 2 MRT VUs doing redundant execution to ensure reliable of 99\%. On the other hand it requires $2100/700 = 3$ checkpoints if each group contains 3 MRT VUs doing redundant execution to ensure reliability of 99\%. The cost paid by VC in first case is 7 checkpoints $\times$ 2 VUs doing redundant execution in single checkpoint $\times$ 300 minutes of execution time in a checkpoint = 4200 unit. In second case, the cost paid by VC is 3 checkpoints $\times$  3 VUs doing redundant execution $\times$ 700 minutes of execution time in a checkpoint = 6300 units. So we can clearly see that task splitting is cost effective to the VC.

It's important to highlight that while task splitting is a feasible option, the split tasks cannot be executed in parallel due to the inherent characteristics of the tasks. Essentially, task splitting functions as a checkpoint. The subsequent split version can only be executed once the preceding split version has been successfully completed.

We define three set of variables: (a) $n_{min}^{s}$ and
$n_{max}^{s}$: representing the minimum and maximum number of SRT VUs that can be allocated for task execution in a single group respectively, (b) $n_{min}^{m}$ and
$n_{max}^{m}$: representing the minimum and maximum number of MRT VUs that can be allocated for task execution in a single group respectively, (c) $n_{min}^{l}$ and
$n_{max}^{l}$: representing the minimum and maximum number of LRT VUs that can be allocated for task execution in a single group respectively. Carefully observing the graph shown in \autoref{fig:MT99R}, we choose the values of the above variables in \autoref{tab:var}. Since the average execution time of the tasks is 1000 minutes with maximum execution time (generated from \cite{googlecluster2019}) being 5000 minutes, we limit the values of the variables. The pseudo-code of the task splitting is present in \autoref{alg:alg_ts}. 

While splitting the tasks the AP try to allocate less number of VUs in a single group starting from $n_{min}^{VU\_type}$ as it is more cost effective to the VC. If allocation is not possible the AP increase the number of allocated VUs in a single group till it reaches $n_{max}^{VU\_type}$. If VU allocation is not possible for that type of VU, then the VC try to allocate different type of VU in order of LRT, MRT and SRT VUs. 

\begin{table}[]
    \footnotesize
    \centering
    \begin{tabular}{|c|c|c|c|c|c|}
    \hline
        $n_{min}^{s}$ & 3 & $n_{min}^{m}$ & 2 & $n_{min}^{l}$ & 2 \\
         \hline
         $n_{max}^{s}$ & 5 & $n_{max}^{m}$ & 4 & $n_{max}^{l}$ & 3 \\
         \hline
    \end{tabular}
    \caption{Values of variables chosen depending on task execution time and $LUT$}
    \label{tab:var}
\end{table}

\begin{algorithm} [tb!]
\footnotesize
    \caption{Task Splitting}
    \label{alg:alg_ts}
    \textbf{Location:} Availability Zone, \textbf{Initiator:} AP of respective VCs\\
    \textbf{Input:} {Task \(\tau_i\) \(\{a_i, e_i, d_i, r_i\}\) \ ,$VU\_type$ = [SRT, MRT, LRT]\\}
    \textbf{Output:} {$g_i \gets$ Number of checkpoints, $VU\_type_i \gets$ Number of VUs of which type allocated for $\tau_i$ execution or message of failed allocation, $x_i \gets$ required number redundant execution}
    \begin{algorithmic}[1]
    
    \FOR{x in range $n^{VU\_type}_{min}$ to $n^{VU\_type}_{max}$}
        \STATE{$g_i \gets \frac{e_i}{MT99R_{T_x}}$}
        \IF{$g_i * x \leq n^{VU\_type}_{VU}$}
        \STATE{Update $g_i$, $x_i$, $VU\_type_i$ for $\tau_i$}
        \STATE{return $val=0$}
        \ENDIF 
        
    \ENDFOR
     \STATE{return $val=-1$}
    \end{algorithmic}
\end{algorithm}

\subsection{Task's VU Requirement Calculation}
The AP accesses the task queue within the availability zone and computes the necessary count of VUs required by each task. It proceeds to accept tasks based on the availability of the required number of VUs. Since the availability zone orders the tasks as decreasing order of priority as decided in \autoref{ss:to}, the AP scans the queue and selects tasks based on their priority if the VC can fulfill the VU requirements of the tasks. It uses the lookup table $LUT$ for calculation. The pseudo code of the approach is shown in \autoref{alg:alg_trq}. While calculating task VU requirement, the AP also calculates efficient task splitting using \autoref{alg:alg_ts}.

Each AP allocates a reserved proportion of 10\% of LRT VUs for high-critical tasks. It calculates its reserved VU number ($T_{VU}$) using its historical LRT VU incoming rate. If the number of idle LRT VUs available in VC ($n^l_{VU}$) is more than $T_{VU}$, the AP prioritises LRT VUs, followed by MRT VUs and SRT VUs for both critical and non-critical tasks. This is because LRT VUs provides more reliability, followed by MRT VUs, followed by SRT VUs. In cases where there is a deficiency of LRT VUs, the AP uses the reserved lot of LRT VUs only for critical tasks, while only MRT and SRT VUs are allocated to non-critical tasks. 
\begin{algorithm} [tb!]
\footnotesize
    \caption{Task's VU Requirement Calculation}
    \label{alg:alg_trq}
    \textbf{Location:} Availability Zone, \textbf{Initiator:} AP of respective VCs\\
    \textbf{Input:} {Task \(\tau_i\) \(\{a_i, e_i, d_i, r_i\}\) \ \\}
    \textbf{Output:}{$g_i$, $x_i$, $VU\_type_i$}
    \begin{algorithmic}[1]
    
    \IF{$\tau_i$ is critical and $n^l_{VU}<T_{VU}$}
        \STATE{$val \gets $Task Splitting ($\tau_i$, $VU\_type$ = LRT) using \autoref{alg:alg_ts} }
        
        \STATE{\textbf{if} $val \neq -1$ return}
    \ENDIF
    \IF{$n^l_{VU}>T_{VU}$}
        \STATE{$val \gets $Task Splitting ($\tau_i$, $VU\_type$ = LRT) using \autoref{alg:alg_ts} }
        
        \STATE{\textbf{if} $val \neq -1$ return}
    \ENDIF
    \STATE{$val \gets $Task Splitting ($\tau_i$, $VU\_type$ = MRT) using \autoref{alg:alg_ts} }
        
        \STATE{\textbf{if} $val \neq -1$ return}
    \STATE{$val \gets $Task Splitting ($\tau_i$, $VU\_type$ = SRT) using \autoref{alg:alg_ts} }
        
        \STATE{\textbf{if} $val \neq -1$ return}

    \end{algorithmic}
\end{algorithm}

\begin{algorithm} [tb!]
\footnotesize
    \caption{VU allocation to task}
    \label{alg:alg_talloc}
    \textbf{Location:} Availability Zone, \textbf{Initiator:} AP of respective VCs\\
    \textbf{Input:} {$\tau_i$} \\
    \textbf{Output:} {The VU allocation for task $\tau_i$}
    \begin{algorithmic}[1]
    \STATE{Calculate $x$, $g_i$ and $VU\_typr$ for $\tau_i$ using \autoref{alg:alg_trq}($\tau_i$)}
    \STATE{Allocate $\tau_i$ with $g_i$ checkpoints and $x$ redundant execution on $VU\_type$ VUs}
    \end{algorithmic}
\end{algorithm}

\subsection{Task Acceptance Procedure}
So far, the AP is aware whether the task is critical or non-critical, estimated the number of VUs needed to execute the task in order to guarantee reliability, and arranged sequence of tasks. The VCs select the tasks depending on some approach as mentioned in \autoref{ss:to}. A task is not chosen by VC, if it can accomodate the number of VUs required. The AP can also discard a task from the sequence if it finds out the task can not be completed within its deadline. \autoref{alg:alg_tap} represents the pseudo-code of this approach. The time complexity of this approach is \(O(N)\), where $N$ is number of tasks at that instant.

\begin{algorithm} [tb!]
\footnotesize
    \caption{Task Acceptance Procedure}
    \label{alg:alg_tap}
    \textbf{Location:} Availability Zone, \textbf{Initiator:} AP of respective VCs\\
    \textbf{Input:} {Task set \textbf{T}=\{${\tau_1,\tau_2,\cdots\,\tau_N}$\}}\\
    \textbf{Output:}{Execution of $\tau_i$ on VUs}
    \begin{algorithmic}[1]
    \FOR{Tasks Availability Zone sequence}
         \STATE Let $\tau_i$ is next task of the set to be considered
        \IF{ $d_i \leq $ $C_t + e_i$}
            \STATE{Reject the task, remove from the task list}
        
        \ELSE
        \IF{Number of VUs present in VC $>n^i_{VU}$}
        \STATE Execute tasks with parameters $g_i$,$n^i_{VU}$ = \autoref{alg:alg_trq}
        \ENDIF
        \ENDIF
    \ENDFOR
    \end{algorithmic}
    
\end{algorithm}

\subsection{Task Migration and VU Allocation}
After AP decides which task to execute, the task migrates to AP inside VC from availability zone. The AP then points out the list of VUs to recruited for task execution from its VU allocation list. The AP selects a collection of VUs in FCFS order, then the task migrates to the collection of VUs.   

\subsection{Task Execution}
After VUs are allocated and the task migrates to allocated VUs task execution starts. Three things can happen while a task is execution: 1. Case A- When a VU or a subset of VUs leave the VC while executing a particular task, then the AP initiates \autoref{alg:alg_VUexit}. 2. CASE B- When all the VUs executing a particular task leaves the VC simultaneously. In that case the AP initiates \autoref{alg:alg_task_fail}. 3. CASE C- When a task is completed, then the AP initiates \autoref{alg:alg_task_complete}. A particular task can experince first two events multiple time, but it experience task completion only once.

\subsection{VU Leaving Handling}
\label{MaxWork}
When a VU departs the VC, the AP finds out due to its maintenance of an up-to-date list of VUs present within the VC at any given time. Once the event of VU departing the VC is detected, it initiates \autoref{alg:alg_VUexit}. The AP assigns some work to recruiter, the pseudo-code of that assigned work is presented in \autoref{alg:alg_ta}. 

The AP on detection of VU leaving the VC while executing task $\tau_i$, it initially checks whether the departing VU is a recruiter. If the leaving VC is a recruiter, then the AP selects another VU as recruiter which is still executing $\tau_i$. The AP calculates if more number of VUs needs to be recruited to ensure reliability, considering the execution time left for task completion. On finding if more VUs are required, it communicates the list of VUs need to be recruited by the recruiter along with information about the departing VU and the list of VUs still executing $\tau_i$.

The recruiter ($VU_R$) on finding $VU_j$ leaving, it first checks if it has executed more amount of task than $VU_j$, if not it updates its task progress with $VU_j$'s task progress. The $VU_R$ continues its task execution as usual and simultaneously checks which VU executing $\tau_i$ has done most progress. On finding that particular VU ($VU_{max}$), it updates the progress to newly recuited VUs and also updates itself if necessary.

\begin{algorithm}[tb!]
\footnotesize
    \caption{VU Leaving Handling}
    \label{alg:alg_VUexit}
    \textbf{Location:} VC , \textbf{Initiator:} AP of VC \\ 
    \textbf{Input:} {List of VUs running task \(\tau_i\)}, $e'_i \gets$ Execution time left for $\tau_i$\\
    \textbf{Output:} {Newly recruited VU group ensuring reliability for $\tau_i$}
    \begin{algorithmic}[1]
    \STATE Let $VU_{leave}$ leaves VC
    \STATE Let $VU_R$ be the recruiter
    \IF{$VU_R$ and $VU_{leave}$ are same vehicle}
        \STATE Select another task executing VU as $VU_R$
    \ENDIF
    
    \STATE $n_{VU} \gets $Calculate number of VUs to be recruited to ensure reliability using $LUT$ with newly left execution time
    \IF{$n_{VU} >$ $n_i$}
        \STATE \autoref{alg:alg_ta} (list of $n_i$VUs executing $\tau_i$, $VU_{leave}$, list of $n_{VU}-n_i$ VUs to be recruited)
    \ENDIF

    \end{algorithmic}
\end{algorithm}

\begin{algorithm}[tb!]
\footnotesize
    \caption{$T_N$ Task Assignment}
    \label{alg:alg_ta}
    \textbf{Location:} VC,  
    \textbf{Initiator:} Recruiter \\ 
    \textbf{Input:} {List of VUs executing \(\tau_i\)}, $VU_{leave}$, {List of VUs to be recruited for execution of \(\tau_i\)}\\
    \textbf{Output:}{Newly recruited VU group ensuring reliability for $\tau_i$}
    \begin{algorithmic}[1]

    \IF{$WorkDone_{VU_{leave}} > WorkDone_{VU_R}$ }
        \STATE $VU_R$ overwrites it's $Workdone$
    \ENDIF
    \STATE $VU_R$ continues task execution
    \STATE $VU_{max}=$ Select the VU executing $\tau_i$ with maximum work done 
     
    \IF{$WorkDone_{VU_{max}} > WorkDone_{VU_R}$}
    \STATE $Workdone_{max}=Workdone_{VU_{max}}$
    \STATE $VU_R$ overwrites it's $Workdone$ with $Workdone_{max}$     
    \ENDIF
    \STATE \textbf{else} $Workdone_{max}=Workdone_{VU_R}$
    \STATE Recruit $n_{VU}-n_i$ VUs
    \STATE Copy newly recruited VUs with $Workdone_{max}$

    \end{algorithmic}
\end{algorithm}

\subsection{Task Failure Procedure}

\par{
The AP continuously monitors the execution of tasks within its VC. It can effectively identify task failures by simply checking if all VUs engaged in executing a particular task depart simultaneously. When a AP detects a task failure, the AP initiates \autoref{alg:alg_task_fail}.

AP simply discard the task for further execution if it finds out it can not be executed within its deadline. But if the deadline is too far away and the revenue given of the task is sufficiently high then the AP re-run the task and earn the profit. So, a new set of VUs are assigned and the task is then re-executed from the beginning, as the progress made in the previous execution is lost. The time complexity of this algorithm is \(O(t_{num})\) where $t_{num}$ is the number of tasks executing in VC.
}
\subsection{Task Completion Algorithm}
The AP periodically monitors the advancement of task execution. Upon detecting task completion it initiates \autoref{alg:alg_task_complete}.

The VC gets the profit of the task, if the task is completed before its deadline. The time complexity of this algorithm is \(O(t_{num})\) where $t_{num}$ is the number of tasks executing in VC.

\begin{algorithm} [tb!]
\footnotesize
    \caption{Task Completion Algorithm}
    \label{alg:alg_task_complete}
    \textbf{Location:} VC, 
    \textbf{Initiator:} AP of VC \\ 
    \textbf{Input:} {Running tasks in VC $T_{t_{num}}$}\\
    \textbf{Output:} {Add task's profit to total profit if completed}
    \begin{algorithmic}[1]
     \STATE $C_t \gets$ current time
     \WHILE{any running task which has not yet considered in current time slot}
     \STATE {let $\tau_i$ is the next considered task}
      \STATE{$F_{work\_completed} = false$}
        \FOR{check all the VUs executing task $\tau_i$}
            \STATE{calculate the work done so far by this VU}
            \IF{$WorkDone=e_i$}
                \STATE{$F_{work\_completed} = true$ 
                \textbf{break loop}}
            \ENDIF
        \ENDFOR
        \IF{$F_{work\_completed} = true$ AND $C_t \leq d_i$}
            \STATE{Add profit of the  task to total profit}
        \ENDIF
            \STATE{Free all the currently assigned VUs for this task $\tau_i$}
            \STATE{Remove this task $\tau_i$ from list of running task}
    \ENDWHILE
    \end{algorithmic}
\end{algorithm}

\begin{algorithm} [tb!]
\footnotesize
    \caption{Task Failure Incorporation}
    \label{alg:alg_task_fail}
    \textbf{Location:} VC, 
    \textbf{Initiator:} AP of VC \\ 
    \textbf{Input:} Running tasks in VC $T_{t_{num}}$\\
    \textbf{Output:} {Add task's profit to total profit if completed}
    \begin{algorithmic}[1]
     \STATE $C_t \gets$ current time 
     \STATE Let $\tau_i$ has failed
     \IF{$C_t+e_i>d_i$}
     \STATE{Remove the task from VC}
    \ELSE
        \STATE{Re-allocate VUs using \autoref{alg:alg_talloc}}
    \ENDIF
     
    \end{algorithmic}
\end{algorithm}

\section{Experimental Results and Analysis}\label{expeval}
\subsection{Simulation Environment}
During our search for simulators capable of accurately representing a real-life parking lots, two options emerged as potential candidates. SUMO \cite{sumo} is a popular simulator in the literature, unfortunately, it is not modelled to simulate a car parking environment. On the other hand, VISSIM \cite{vissim} does offer the capacity to simulate parking lots. However, due to challenges in obtaining access to VISSIM, we decided on not using it. So, we decided to develop a custom simulator using C++ programming to address the specific requirements of this parking-related problem.

\subsection{State of the Art Approach}

We consider the approach proposed by Florin \etal \cite{Florin21} as the state-of-the-approach (SOTA). Our approach enhances the SOTA approach in the following ways:

\begin{enumerate}
    \item In the SOTA, all VUs are considered identical in terms of residency time, while our approach categorizes VUs into Short Residency Time (SRT), Medium Residency Time (MRT), and Large Residency Time (LRT) based on their residency durations. Consequently, each VU's predicted residency duration becomes distinct as mentioned in \autoref{RATime}.
    \item The previous approach did not classify tasks into critical and non-critical types , whereas our approach includes such classification and presents different methods to handle them. 
    \item In the SOTA , the task is not split into multiple checkpoints, where as in our proposed approach split the tasks into multiple checkpoints, which effects in reduction of VU computation resource wastage. 
    \item  In the SOTA approach, while performing \(T_n\) task assignments,the recruiter replicates the VM image of the departing VU to the next recruited VU. Conversely, in our approach, we replicate the VM image of the VU that has done the maximum amount of work to the newly recruited VU, as mentioned in \autoref{jntaskass}. Moreover, our approach of $T_n$ task assignment also considers execution time of the task left till completion in case of recruiting VU after a VU leaves, which effectively increase the profit of the VC by reducing cost of task execution and increase number of task execution.
    
\end{enumerate}

\subsection{Dataset Generation}
\begin{enumerate}
    \item VU Dataset: The arrival time of each VU follows Poisson distribution and the execution time follows exponential distribution. We considered $70\%$ of VUs have small residency period with $t_{mrt}^s$ to be $120$ minutes. Around $20\%$ of VUs have medium residency period with $t_{mrt}^m$ to be $200$ minutues. And the remaining $10\%$ VUs have large residency period with $t_{mrt}^l$ to be $400$ minutes. We have also incorporated a uncertainty in 10\% VUs where the VU leaving time is much different than previously predicted.
    \item Task Dataset: The arrival time of each incoming task follows Poisson distribution and the execution time follows exponential distribution. Additionally the laxity of a task $l_i$ is 10\% of the execution time along with a randomly generated value that ranges from 100 minutes to 5000 minutes. And henceforth the deadline $d_i$ of task $\tau_i$ is given by $d_i = a_i + e_i + l_i$. And the price of task $\tau_i$ which the VC receives on successful completion of task is dependent on the execution time and the closeness of its deadline and is defined as:
    \begin{equation}\label{}
        p_i = K_1{e_i}^{1.5} + \frac{K_2}{{l_i}^{2}}
    \end{equation}
    Apart from this we have also included some outliers tasks which have large execution times as nothing in the world strictly follows a particular distribution and is completely symmetrical.
\end{enumerate}

\subsection{Performance with Variation of Parameters}
In this sub-section we analysis the performance variations of different approaches as the hyper parameters change. As a continuation we also see the performance of the State of The Art approach (SOTA) versus our proposed approach. The base values for all the hyper parameters are mentioned below:

\begin{table}[tb!]
\centering
\footnotesize
\begin{tabular}{ | l | l | } 
  \hline
  \textbf{Parameters} & \textbf{Base values}  \\
  \hline
  Number of tasks & 1000\\ \hline
  Number of VUs & 10000\\ \hline
  \(t_{mrt}^s\), \(t_{mrt}^m\), \(t_{mrt}^l\) & $120$,  $200$,  $400$ minutes\\ \hline
  Outliers for each VU type & 10 \%\\ \hline
  \(k_1\), \(k_2\) & 3, 10000\\   \hline
\end{tabular}
\caption{Base values used for the experiments}
\label{Table:Hyperparam}
\end{table}

\subsubsection{Performance of Different Task Ordering Approaches versus Number of Tasks}
\begin{figure*}[tb!]
\centering
\begin{subfigure}{0.48\textwidth}
    \includegraphics[width=\textwidth]{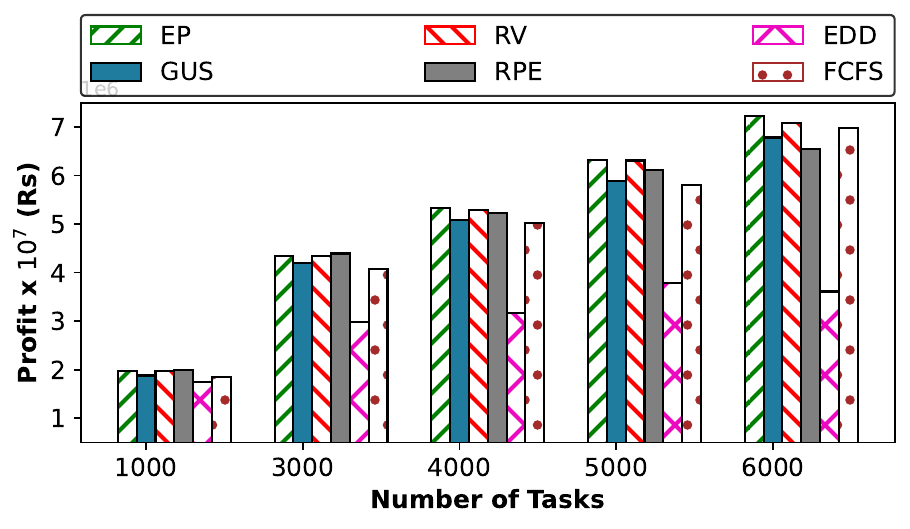}
    \caption{Profits Vs Number of Tasks (Number of VUs = 10000)}
    \label{fig:task_var}
\end{subfigure}
\hfill
\vspace{0.5 cm}
\begin{subfigure}{0.48\textwidth}
    \includegraphics[width=\textwidth]{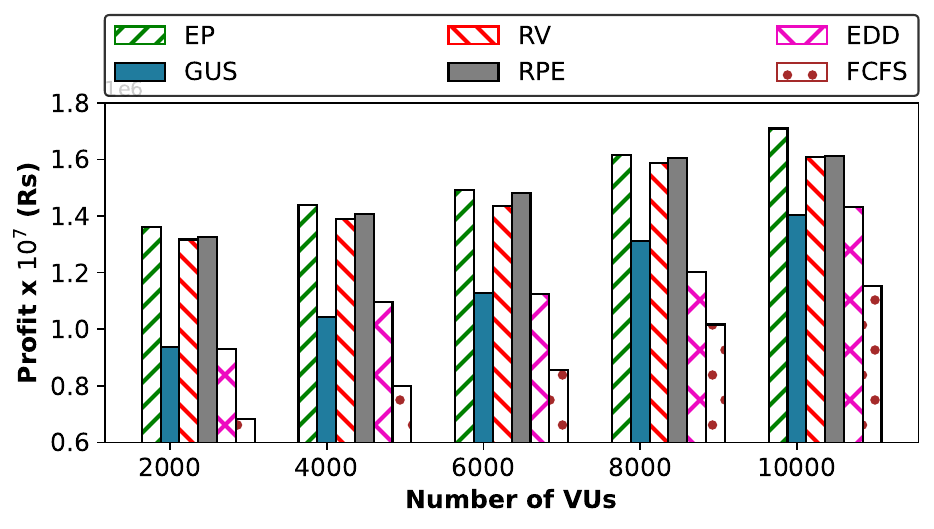}
    \caption{Profits Vs Number of VUs (Number of tasks = 1000)}
    \label{fig:car_var}
\end{subfigure}
\hfill
\begin{subfigure}{0.48\textwidth}
    \includegraphics[width=\textwidth]{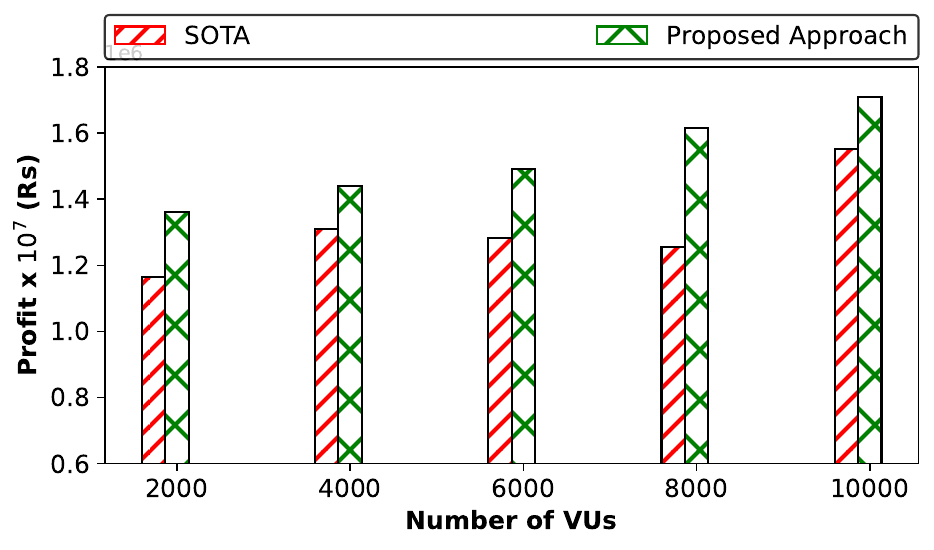}
    \caption{Profit Vs Number of VUs (Number of tasks = 1000)}
    \label{fig:us_v_sota_car}
\end{subfigure}
\hfill
\begin{subfigure}{0.48\textwidth}
    \includegraphics[width=\textwidth]{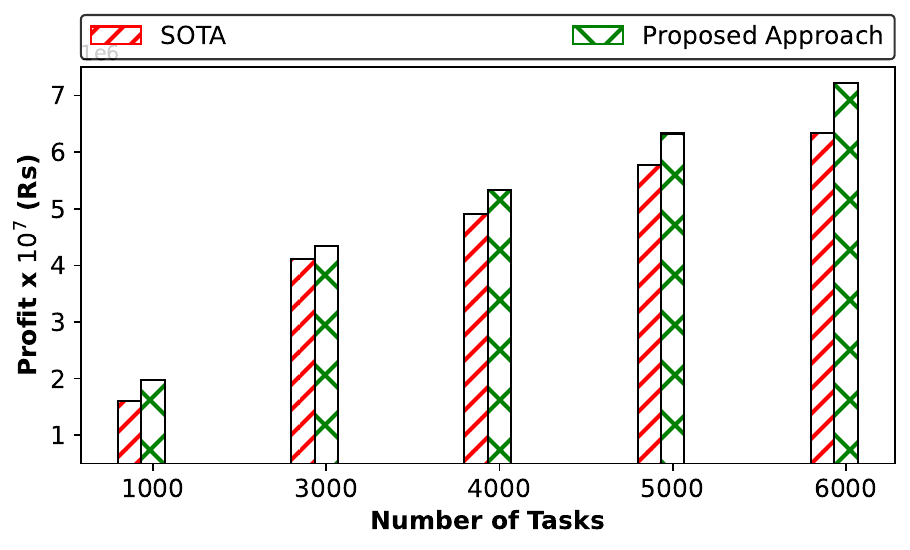}

    \caption{Profit Vs Number of Tasks (Number of VUs = 10000)}
    \label{fig:us_v_sota_task}
\end{subfigure}
\vspace{0.5 cm}
\caption{Sensitivity analysis for number of tasks and VUs}
\label{fig:figures}
\end{figure*}
\autoref{fig:task_var} shows the profit earned by different task ordering approaches as number of tasks vary. We see the performance of different scheduling approaches with the variation of the number of tasks given the number of VUs is fixed at 10,000 in all the cases. As the number of tasks increase from 1000 to 3000 then we can see a significant improvement in the profit. This is because many of the VUs were getting unused in the case of 1000 tasks and hence we had surplus of VUs (because when many VUs are available then it might happen that no task is present), and hence forth we were able to serve 3000 tasks as well which in turn resulted in an increase of our profit. The profits increase further because the VC can choose higher profit tasks when the number of tasks is more. 

EP in most of the case outperforms the other approach because it also accounts for the cost of the task while scheduling. Approaches like RV does not consider the length of the task to be executed in any way, hence the cost is not taken into account resulting into slightly poor performance. Approaches like EDD and FCFS do not not even consider the revenue of the task while scheduling, resulting into very bad performance. 
\subsubsection{Performance of Different Task Ordering Approaches versus Number of VUs}
\autoref{fig:car_var} shows the performance of different scheduling approaches as the number of VUs vary, when the number tasks are constant. Here the number of task is fixed at 1000 with mean execution time of 1000 minutes. Clearly as the number of VUs assigned to execute the tasks increase, the count of successfully completed tasks also experience an upward trend. So, the total profit of the VC increases. Another factor that contributes the increase in profits is the VC do not need to pay the VUs when they remain within the VC premises without actively executing any tasks. So, the VC cost is not effected by idle VUs.

As we can see that EP outperforms all other approaches both when number of VUs and tasks increase, so we have considered it to be our primary task scheduling approach in all the later experiments going forward.
\subsubsection{Performance of the Proposed Approach versus State of The Art Approach with varying Number of VUs and Number of Tasks}

We clearly see in \autoref{fig:us_v_sota_car}, \autoref{fig:us_v_sota_task} that our proposed algorithm beats the state of the art (SOTA) approach significantly in profit collection with EP as the scheduling approach. The profits increase further as the number of tasks, VUs increase. The reason being is that SOTA does not differentiate among different types of VUs in the parking lot and does not take advantage of the varying residency times of VUs hence large VUs are also getting allocated along with small and medium VUs. Moreover, we split the longer tasks into shorter parts and create checkpoints, to reduce VU computation wastage. Also one more major difference is that when a VU leaves the mall and recruiter recruits a new VU, in case of SOTA the amount of work done by the leaving VU is getting copied to the newly allocated VU whereas in proposed algorithm the maximum work done till now is getting copied to the newly allocated VU.

\subsubsection{Effect of variation in Mean Execution Time and Mean Residency Time on Profit Percentage of Proposed Approach}

\begin{figure}[tb!]
\centering
    \includegraphics[width=0.5\textwidth]{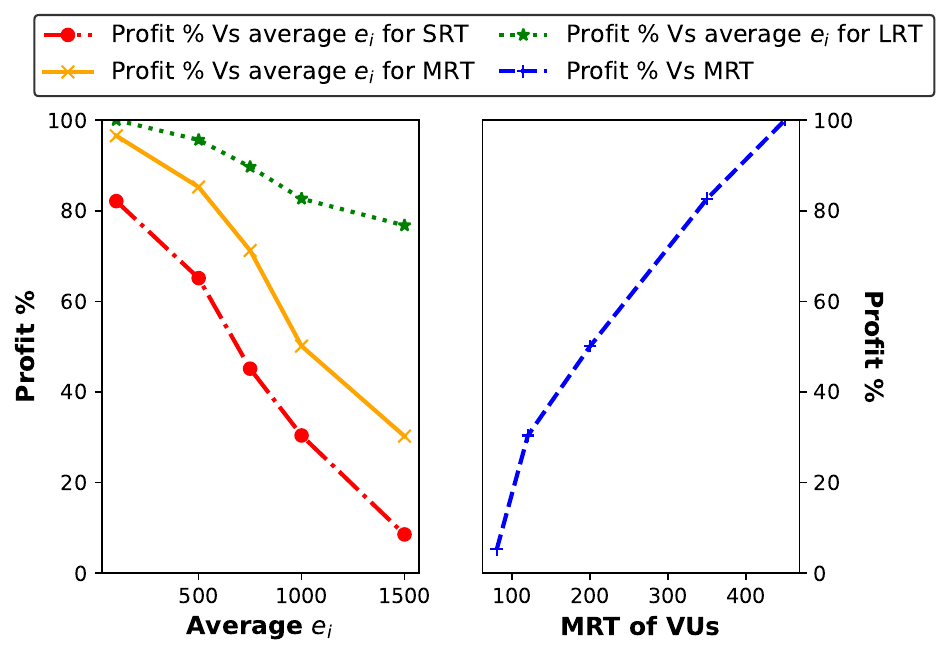}
    \caption{Profit (\%) with EP scheduling Vs Variation of mean execution time (left) and MRT (right), number of VUs = 10000}
    \label{fig:line}
\end{figure}

We can see from \autoref{fig:line} that the profit percentage reduces as the mean execution time of tasks increase. This is because the increase in task execution time increases the chances of failure. We also observe that profit percentage is higher in LRT than MRT and SRT VUs respectively. This finding aligns with the findings of \autoref{fig:MT99R} and \autoref{fig:probab}. Again we notice that rate of fall of profit percentage is lower in LRT than MRT and SRT respectively. This is because time ensuring $MT99R$ increases exponentially with MRT. 

Furthermore, the exponential growth of profit percentage with respect to MRT of VUs further validates the observation made in \autoref{fig:probab}. This outcome is rooted in the fact that higher VU residency times correspond to a decreased probability of task failure. Overall, these trends highlight the intricate relationship between task execution times, VU residency times, and profit percentages in our approach.

\subsection{Evaluation using Real Life Dataset}


We take into account two real-world datasets for the amount of time that VUs stay in parking spaces, namely those from Queens \cite{dataset_grand_arcade} and the Grand Arcade shopping mall \cite{dataset_queen_annes_terrace}. The data mainly consists of the number of VUs for different stay periods. Along with this we use data world \cite{dataset2023} to get the arrival in order to generate real life traces. For tasks we have used \cite{googlecluster2019} and scaled up the mean execution time to 1000 minutes.

\subsubsection{Case Study 1: Queens Anne's Terrace Parking}
It is located at Gonville Pl, Cambridge CB1 1ND, USA, at 0.134 longitude and 52.2009 latitude. The VU stay data is for the first week of April 2012, May 2012, June 2012. \\

\begin{figure}[tb!]
\hspace{-20 pt}
\centering
    \includegraphics[width=0.47\textwidth]{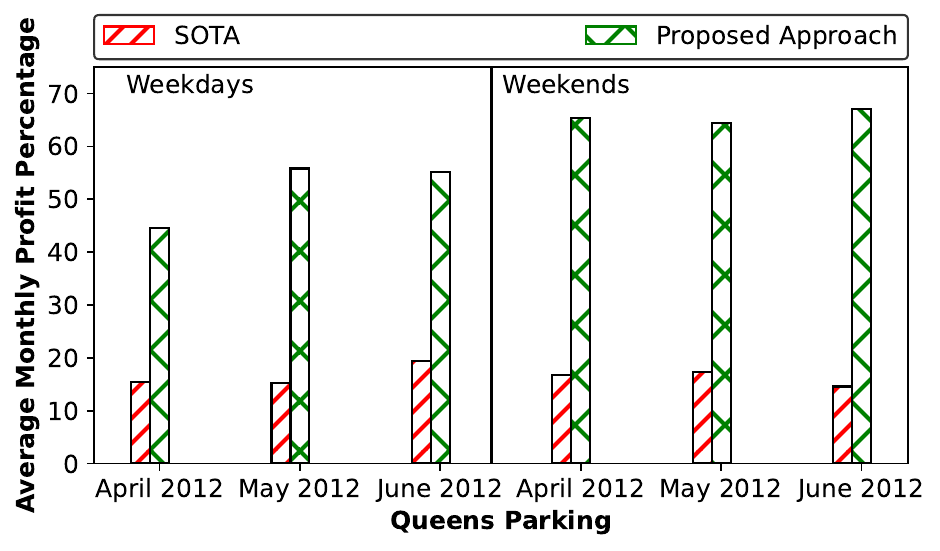}
    \caption{Profit for Queens Anne's Terrace Parking Dataset}
    \label{fig:real_q}
\end{figure}

As evident from \autoref{fig:real_q}, our approach significantly outperforms the SOTA approach in all the real-life traces. The reason for this notable enhancement is our technique is incorporated with multiple uncertainty handling techniques. Additionally, we observe that the profit percentage increase during the weekends in comparison to the weekdays. This is because the number of VUs and their average stay duration is higher in weekends. This observation aligns with our initial findings from \autoref{fig:us_v_sota_car} and \autoref{fig:line} , which demonstrate a positive correlation between VU profit and the number of VUs as well as their mean residency time.

\subsubsection{Case Study 2: Grand Arcade Parking}
\begin{figure}[!]
\hspace{-20 pt}
\centering
    \includegraphics[width=0.47\textwidth]{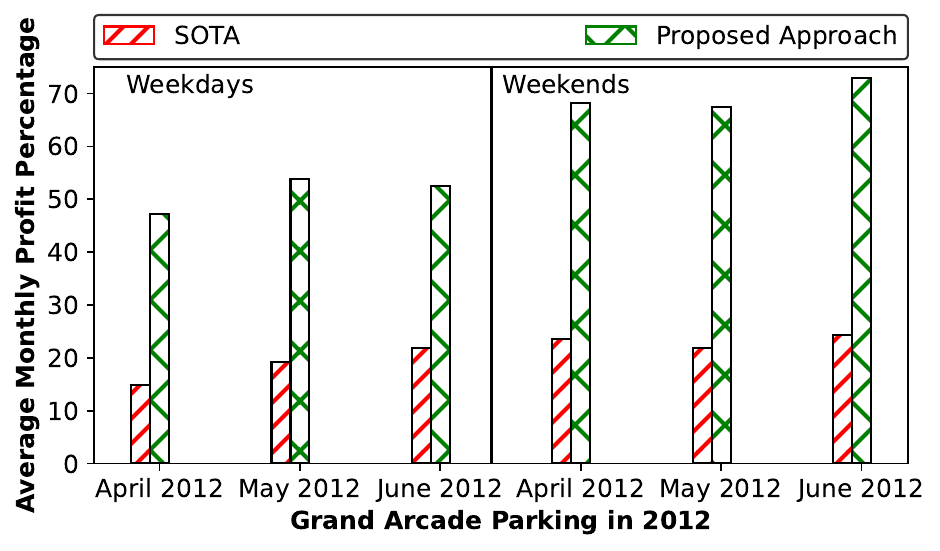}
    \caption{Profit  for Grand Arcade Parking Dataset}
    \label{fig:real_ga}
\end{figure}
It is located at St Andrew's St, Cambridge CB2 3BJ, USA, at 0.1216 longitude and 52.204 latitude. The VU stay data is for the first week of April 2012, May 2012, June 2012.\\

We can see from \autoref{fig:real_ga}, Grand Arcade parking also exhibits similar results to Queens Anne's terrace parking. Our approach outperforms the SOTA by a big margin and also the profit percentage is higher in weekend than in weekdays.

\section{Conclusion and Future Work}
\label{concl}

In this paper, we have tackled an emerging challenge in the field of vehicular cloud systems, contributing a heuristic solution to a realistic problem that has remained unaddressed until now. Our focus on the static vehicular cloud paradigm has enabled us to shed light on the importance of financial considerations and the integration of high reliability into this context – factors that are vital for the viability of such systems. The outcomes of our approach is promising, revealing a substantial 200\% increase in profit percentage compared to state of the art approach in real life dataset.

Looking ahead, there are several modifications that we have not considered. Such modifications includes- VUs providing heterogeneous services, tasks require multiple types of service from the vehicular cloud. Additionally, the development of a mathematical model based on our approach could provide a formalized framework for analysis and prediction, contributing to the broader field of vehicular cloud optimization.  

By unveiling new dimensions and achieving notable improvements, our work serves as a stepping stone towards a more efficient, reliable, and profitable vehicular cloud ecosystem.

\setstretch{0.7}
\bibliographystyle{unsrt}
\footnotesize
\bibliography{main}
\end{document}